\documentstyle[preprint,prd,aps,epsfig]{revtex}
\tightenlines
0
\begin{document}
\draft

\title{Analytic Confinement and Regge Trajectories}

\author{G.V.~Efimov and G.~Ganbold\footnote{{\it Permanent address:}
Institute of Physics and Technology, Mongolian Academy of
Sciences, \\ 210651 Ulaanbaatar, Mongolia}}

\address{Bogoliubov Laboratory of Theoretical Physics, \\
Joint Institute for Nuclear Research, 141980 Dubna, Russia}

\date{\today}

\maketitle

\begin{abstract}
A simple relativistic quantum field model with the Yukawa-type
interaction is considered to demonstrate that the analytic
confinement of the constituent ("quarks") and carrier ("gluons")
particles explains qualitatively the basic dynamical properties
of the spectrum of mesons considered as two-particle stable bound
states of quarks and gluons: the quarks and gluons are confined,
the glueballs represent bound states of massless gluons, the masses
of mesons are larger than the sum of the constituent quark masses
and the Regge trajectories of mesonic orbital excitations are almost
linear.
\end{abstract}

\vskip 1cm


\section{Introduction}

The meson spectroscopy as the theory of bound states of quarks, and
the phenomenology of the Regge trajectories (RTs) are important
and interdependent subjects of investigation in particle physics
\cite{regg59,chew62,coll77}.

At the present time, QCD is commonly considered the true theory of
the strong interaction describing all processes in the hadron world,
including the mesonic spectroscopy \cite{wein95}. However, being a
nonlinear theory with local colour gauge symmetry, QCD is quite
complicated from the computational point of view and corresponding
methods of calculations require great efforts in additional
assumptions and ideas. In contrast to QED, simple and reliable methods
of calculations are still missing in QCD. From our point of view, the
main puzzle is that any acceptable and well established description of
quarks and gluons explaining the hadronization on large distances,
where the confinement of quarks and gluons takes place, has not been
found yet. On the other hand, one may expect that a theoretical
description of colorless hadrons considered as bound states of quarks
and gluons, when the confinement is taken into account and the
averaging over all non-observable color degrees of freedom is
performed, can lead to a physical picture, where the quarks and gluons
are realized in the form of some phenomenological "bricks". We suppose
that a successful guess of the structure of these "bricks" in the
confinement region can result, particularly, in a qualitatively
correct description of the basic features of the meson spectrum. Our
guess is that the analytic confinement realizes these "bricks".

Following this idea, in the present paper we consider the meson
spectroscopy within the theory of bound states of quarks and gluons.
Our aim is to show within a simple relativistic quantum field model
that the basic features of experimentally observed meson spectrum
(see, e.g. \cite{tang00}) can be qualitatively explained by the
analytic confinement of quarks and gluons. The basic dynamic
characteristics of spectrum of all mesons considered bound states of
quarks and gluons (in contrast to the relations of the $SU_3$ flavor
symmetry) can be listed as follows:
\begin{quote}
- quarks and gluons are confined;

- glueballs are bound states of massless gluons, i.e. they are
  completely relativistic systems;

- the masses of mesons are larger than the sum of masses of the
  constituent quarks;

- the RTs of different families of mesonic orbital excitations
  are quite close to linear and their slopes practically coincide.
  It means that the slope is an universal parameter which is defined
  by   the general nature of quark-gluon interaction.
\end{quote}

Obviously, these characteristics cannot be obtained in the framework
of any local quantum field theory, where the constituent particles,
quarks and gluons, are described by standard Dirac and Klein-Gordon
equations. From the common point of view the confinement plays the
first role in understanding and explaining of this picture. The point
is how to realize mathematically the conception of confinement in
specific theoretical formalism?

The standard theoretical calculations leading to linear RTs of mesonic
bound states of quarks and gluons are based on the following
assumptions. First, the quarks are accepted as ordinary fermions
interacting by means of gluons. Some argumentations including lattice
calculations of the nonlinear QCD gluon dynamics are given to get a
particular infrared behavior of the gluon propagator
($\tilde{D}(p^2)\sim{1/p^4}$ for $p^2\to 0$) which results in a linear
increasing potential between quarks in three-dimensional space
${\bf x} \in {\bf R}^3$  (see, for example, \cite{godf85}). Exactly
this infrared behavior is interpreted as the quark confinement.

Second, one of three-dimensional reductions of the relativistic
Bethe-Salpeter equation is used. For this purpose it is necessary to
overcome some mathematical problems caused by the singularity of the
gluon kernel and an ambigously defined choice of particular reduction
of the relativistic two-body Bethe-Salpeter equation (see, for example,
\cite{mau93}). As a result a relativized Scr\"{o}dinger equation
is obtained in ${\bf R}^3$-space to estimate the meson masses
including higher orbital excitations. Hereby, the kinetic energy of
the quarks looks like $E_{kin}=\sqrt{{\bf p}^2+m^2}$  and the
potential energy increases linearly. The most essential part of the
two-body Hamiltonian takes the form
$$
H = \sqrt{{\bf p}_1^2+m_1^2} + \sqrt{{\bf p}_2^2 + m_2^2}
+ \alpha|{\bf r}_1-{\bf r}_2| \,.
$$

In the present paper we demonstrate another possible mechanism
explaining the above mentioned characteristics of mesonic spectrum,
including the origin of linear RTs. In doing so, we use a simple
relativistic quantum-field model of two scalar particles (the
prototypes of constituent "quarks" and intermediate "gluons") with
the analytic confinement. Our approach is based on the following
assumptions:
\begin{quote}
- the analytic confinement takes place, it means that propagators
  of "quarks" and "gluons" are entire analytic functions of the first
  order on the complex momentum $p^2$-plane;

- the interaction is described by the Yukawa type Lagrangian;

- the coupling constant binding the "quarks" with "gluons" is small;

- final bound states of "quarks" are described by the relativistic
  Bethe-Salpeter equation in one-"gluon" exchange without using any
  three-dimensional reduction.
\end{quote}

In the present paper we shall not discuss the origin and details of
the analytic confinement. Particularly, the self-dual homogeneous
vacuum gluon field in QCD results in the analytic confinement of
quarks and gluons \cite{leut81,efim93}. The model of induced quark
currents based on this vacuum gluon field describes correctly the
experimental data \cite{efned99}.

This paper is aimed to demonstrate the mathematical sketch of
calculations of the two-body bound-state spectrum within the
Bethe-Salpeter equation within simple QFT models, when the analytic
confinement takes place in the weak-coupling regime. These simple
relativistic models are based on physically transparent hypotheses
and can be treated by simple analytic methods. We claim that the
analytic confinement is the basic underlying "brick" leading to the
qualitatively correct description of main characteristics of the meson
spectra. In any case we believe that our models represent certain
theoretical interest because they clarify the underlying physical
principles of the meson spectrum.

For simplicity, we deal with two possible versions of the analytic
confinement. In the first model the propagators of "quarks" and
"gluons" are pure Gaussian exponents, i.e.
$\tilde{D}(p^2)\sim e^{-{p^2/\Lambda^2}}$, where the parameter
${1/\Lambda}$ represents the "radius" of confinement. From physical
point of view this model is important because the eigenfunctions and
eigenvalues of the relativistic Bethe-Salpeter equation within
one-particle exchange approximation can be found explicitly and the
obtained RTs are purely linear. One can say that this model can be
considered a "relativistic oscillator", i.e. it is a relativistic
generalization of the quantum nonrelativistic oscillator with
equidistant spectra.

The second model comes from a natural physical picture of standard
particles with a relevant scale of mass $m$. It is supposed that the
analytic confinement characterized by the scale $\Lambda$ makes
the propagators of particles ("quarks" and "gluons") entire analytic
functions. As a result there are two physical parameters: the scale
of confinement $\Lambda$ and the mass of quark $m$. An interesting
problem is to investigate the dependence of the meson spectrum on the
relative parameter $\mu=m/\Lambda$. We shall show that this model
describes qualitatively well all above listed dynamical
characteristics of meson spectra.

\section{Two-particle Bound States}

We consider the Yukawa model of two interacting scalar fields
$\Phi(x)$ and $\varphi(x)$ described by the following Lagrangian in
the Euclidean space-time as follows
\begin{eqnarray}
\label{Lagran}
{\cal L}(x)=-\Phi^+(x)S^{-1}(\Box)\Phi(x)
-{1\over2}\varphi(x)D^{-1}(\Box)\varphi(x)-g\Phi^+(x)\Phi(x)\varphi(x).
\end{eqnarray}
The coupling constant $g$ is supposed to be sufficiently small.

We postulate that the analytic confinement takes place here. It means,
that propagators $\tilde{S}(p^2)$ and $\tilde{D}(p^2)$ of confined
particles $\Phi$ and $\varphi$ are entire analytic functions in the
complex $p^2$-plane so that the functions $S^{-1}(z)$ and $D^{-1}(z)$
have no zeros at any finite complex $z$. As a result the equations
for free fields
\begin{eqnarray}
\label{fr-eq}
S^{-1}(\Box)\Phi(x)=0,~~~~~~~~D^{-1}(\Box)\varphi(x)=0
\end{eqnarray}
result only in the unique trivial solutions $\Phi(x)\equiv 0$ and
$\varphi(x)\equiv 0$. We call this property {\it analytic confinement},
i.e. the corresponding particles exist only in virtual states, so
$\Phi$ and $\varphi$ are the {\it virton fields} (see \cite{efim93}).
One can say that these fields describe constituent particles, e.g.
$\Phi(x)$ and $\varphi(x)$  represent  scalar "quarks" and scalar
"gluons", correspondingly.

As mentioned above, we deal with two simple versions of the analytic
confinement.

First, we consider the simplest, but important case with pure Gaussian
propagators
\begin{eqnarray}
\label{prop1}
&& \tilde{S}(p^2)={\epsilon\over\Lambda^2}\,e^{-{p^2\over\Lambda^2}},
\qquad S(x)={\Lambda^2\epsilon\over(4\pi)^2}
\cdot e^{-{1\over4}\Lambda^2 x^2}, \\
&& \tilde{D}(p^2)={1\over\Lambda^2}\cdot e^{-{p^2\over\Lambda^2}},
~~~~~~~~~~D(x)={\Lambda^2\over(4\pi)^2}\cdot
e^{-{1\over 4}\Lambda^2 x^2},\nonumber
\end{eqnarray}
where the parameter $\epsilon\ll 1$ implies that $\Phi$-particle is
much "heavier" than $\varphi$-particle. In some sense this model can
be considered a relativistic "oscillator" because the corresponding
Bethe-Salpeter equation can be solved exactly and the exact solution
gives linear RTs. We call this case {\it the virton model}.

The second model implies that there exists a certain physical
mechanism which generates analytic confinement of standard particles
with the initial masses $m$ and $0$. Then, their propagators are given
in more realistic forms
\begin{eqnarray}
\label{prop2}
&& \tilde{S}(p^2)={1\over p^2+m^2}\cdot
\left(1-e^{-{p^2+m^2\over\Lambda^2}}\right),~~~~~
S(x)={\Lambda^2\over(4\pi)^2}\int\limits_0^1{d\alpha\over\alpha^2}
e^{-{m^2\over\Lambda^2}\alpha-{\Lambda^2x^2\over4\alpha}},\nonumber \\
&& \tilde{D}(p^2)={1\over p^2}\cdot
\left( 1-e^{-{p^2\over\Lambda^2}} \right),~~~~~~~~~~~
D(x) = {1\over(2\pi)^2x^2}e^{-{\Lambda^2 x^2 \over 4}}\,.
\end{eqnarray}
The confinement region is characterized by the parameter $\Lambda$.
In the deconfinement limit $\Lambda\to0$ one obtains the standard
propagators of massive and massless scalar particles. Within this
model we want to analyze the influence of the parameter $\mu$ on the
behavior of the meson spectrum. We call this case {\it the scalar
confinement model}.

"Two-quark" bound states can be found in the following way. Let us
consider the partition function
\begin{eqnarray}
\label{Z}
&& Z=\int\!\!\!\int\!\!\!\int\delta\Phi\delta\Phi^+\delta\phi~
e^{-(\Phi^+S^{-1}\Phi)-{1\over2}(\varphi D^{-1}\varphi)
-g(\Phi^+\Phi\varphi)}\,.
\end{eqnarray}
By integrating over $\varphi$ we arrive at
\begin{eqnarray}
\label{ZZ}
&& Z=\int\!\!\!\int\delta\Phi\delta\Phi^+~
e^{-(\Phi^+S^{-1}\Phi)+{g^2\over2}(\Phi^+\Phi D\Phi^+\Phi)}.
\end{eqnarray}
The term $L_2[\Phi]=(\Phi^+\Phi D\Phi^+\Phi)$ can be transformed as
\begin{eqnarray*}
L_2[\Phi]&=&{g^2\over2}\int\!\!\!\int dx_1dx_2~\Phi^+(x_1)\Phi(x_1)
D(x_1-x_2)\Phi^+(x_2)\Phi(x_2)\\
&=&{g^2\over2}\int dx\int\!\!\!\int dy_1dy_2~\sqrt{D(y_1)}J(x,y_1)
\delta(y_1-y_2)\sqrt{D(y_2)}J^+(x,y_2)\,,
\end{eqnarray*}
where $x_1=x+{y\over2}$, $x_2=x-{y\over2}$ and
\begin{eqnarray*}
&& J(x,y)=\Phi^+\left(x+{1\over2}y\right)\Phi\left(x-{1\over2}y\right)
=\Phi^+(x)e^{{y\over2}\stackrel{\leftrightarrow}{\partial}}\Phi(x),\\
&&~~~~~~~~~J^+(x,y)=J(x,-y).
\end{eqnarray*}

Let us introduce an entire orthonormal system $\{U_Q(y)\}$, where
$Q=\{n,l,\{\mu\}\}$ may be considered as a set of radial $n$, orbital
$l$ and magnetic $\{\mu\}=(\mu_1,...,\mu_l)$ quantum numbers. We have
the conditions
\begin{eqnarray*}
\int dy~U_{Q}(y)U_{Q'}(y)=\delta_{QQ'},~~~~~~~
\sum\limits_{Q}U_{Q}(y)U_{Q'}(y') = \delta(y-y').
\end{eqnarray*}
Then,
\begin{eqnarray*}
&& L_2[\Phi]={g^2\over2}\sum\limits_{Q}\int dx~J_Q(x)\cdot J_Q(x),
\qquad J_Q(x)=\Phi^+(x)V_Q(\stackrel{\leftrightarrow}{\partial})
\Phi(x), \\
&& J_Q^+(x)=J_Q(x) \,,\qquad V_Q(\stackrel{\leftrightarrow}{\partial})
=i^l\int dy~\sqrt{D(y_1)} U_Q(y)
e^{{y\over2}\stackrel{\leftrightarrow}{\partial}}.
\end{eqnarray*}

By using the Gaussian functional representation we write
\begin{eqnarray*}
e^{L_2[\Phi]}
&=&e^{{g^2\over2}\sum\limits_{Q}\int dx~J_Q(x)\cdot J_Q(x)}  \\
&=& \int\prod\limits_Q\delta B_Q~e^{-{1\over2}\sum\limits_{Q}(B_Q B_Q)
+g\sum\limits_{Q}\int dx~B_Q(x)J_Q(x)} \,.
\end{eqnarray*}

We substitute this representation into the partition function
(\ref{ZZ}) and integrate over $\Phi$. The result reads
\begin{eqnarray}
\label{ZZZ}
Z&=&
\int\prod\limits_Q\delta B_Q~e^{-{1\over2}\sum\limits_{Q}(B_Q B_Q)
-{\rm Tr}\ln(1-gB_QV_QS)}   \nonumber\\
&=&\int\prod\limits_Q\delta B_Q~e^{-{1\over2}\sum\limits_{Q Q'}
(B_Q[\delta_{QQ'}-\lambda\Pi_{Q Q'}]B_{Q'})+W_I[B]}.
\end{eqnarray}
Here
$$ W_I[B]=-{\rm Tr}\left[\ln(1-gB_QV_QS)+
{g^2\over2}B_Q V_QSB_{Q'}V_{Q'}S\right] $$
is a functional, which describes the interactions of the fields $B_Q$.

The polarization operator $\lambda\Pi_{QQ'}$ in the one-loop
approximation is
\begin{eqnarray}
\label{pol}
\lambda\Pi_{QQ'}(z)&=& g^2 \, {\rm Tr}(V_Q~S~V_{Q'}~S)  \\
&=& \int\!\!\!\!\int dy_1 dy_2~U_Q(y_1)\lambda
\Pi(z;y_1,y_2)U_{Q'}(y_2) \,, \nonumber\\
\lambda\Pi(z;y_1,y_2)
&=& g^2 \, \sqrt{D(y_1)} \, S \, \left(z+{y_1-y_2\over2}\right)
S \left(z-{y_1-y_2\over2}\right)\sqrt{D(y_2)}  \,,  \nonumber
\end{eqnarray}
where $z=x_1-x_2$ and $\lambda=\left({g/ 4\pi\Lambda}\right)^2$.
Its Fourier transform reads
\begin{eqnarray}
\label{kernel}
&& \lambda\tilde{\Pi}_p(y_1,y_2)\\
&&= g^2\sqrt{D(y_1)}
\int dz~e^{ipz}S\left(z+{y_1-y_2 \over 2}\right)
S\left(z-{y_1-y_2\over 2}\right)\sqrt{D(y_2)} \,. \nonumber
\end{eqnarray}

The orthonormal system of functions $\{U_{Q}(y)\}$ should
diagonalize the kernel (\ref{kernel}). For this aim we
consider the eigenvalue problem
\begin{eqnarray}
\label{eigen}
\int dy'~\lambda\tilde{\Pi}_p(y,y')U_{Q}(y)=E_Q(-p^2)U_{Q}(y),
\end{eqnarray}
where $E_Q(-p^2)=E_{nl}(-p^2)$, i.e. the eigenvalues are degenerated
over the magnetic quantum numbers $\{\mu\}$.

If the functions $\{U_{Q}(y)\}$ are found, the polarization operator
$\tilde{\Pi}_p(x,y)$ is diagonal on these eigenfunctions
\begin{eqnarray}
\label{diag}
\lambda\tilde{\Pi}_p(y,y')=\sum\limits_{Q} E_Q(-p^2)U_Q(y)U_Q(y')\,.
\end{eqnarray}

We note that the diagonalization of $\tilde{\Pi}_p(y,y')$ is nothing
else but the solution of the Bethe-Salpeter equation in the one-boson
exchange approximation. To get the standard form of the Bethe-Salpeter
equation, we have to introduce new functions
$U_Q(y)=\sqrt{D(y)}\Psi_Q(y)$ (see, e.g., \cite{efim99}) and go to the
momentum space.

So, the partition function (\ref{ZZZ}) becomes
\begin{eqnarray}
\label{Z5}
Z&=&\int \prod\limits_{Q}\delta \tilde{B}_Q~
e^{-{1\over2}\sum\limits_{Q}(B_QG_Q^{-1}B_Q)+W_I[gB]},\\
&& G_Q^{-1}=G_Q^{-1}(-\Box)=1-E_Q(\Box),~~~~~~~p^2=-\Box. \nonumber
\end{eqnarray}
The functional integral (\ref{Z5}) is given over a Gaussian measure
defined by the operator $G_Q^{-1}(-\Box)$.

We would like to stress that this representation is completely
equivalent to the initial one (\ref{Z}). From physical point of view,
we pass on from the world containing fields $\Phi$ and $\phi$ to the
world of bound states $\{B_Q\}$.

The fields $\{B_Q\}$ can be interpreted as the fields of particles
with quantum numbers $Q$ and masses $M_Q$, if the Green functions
$G_Q(p^2)$ have simple poles in the Minkowski space ($p^2=-M_Q^2$),
in other words the equations
\begin{eqnarray}
\label{mass}
&& 1=E_Q(M_Q^2)
\end{eqnarray}
define the masses $M_Q=M_{nl}$ of two-particle bound states with
quantum numbers $Q=(nl)$. These states are degenerated over magnetic
quantum numbers $\{\mu\}$. The operator $G_Q^{-1}(-\Box)$ defines the
kinetic term of the field $B_Q$. To represent this operator in the
standard form we expand it in the vicinity of $p^2=-M_Q^2$ as follows
\begin{eqnarray*}
1-E_Q(-p^2) = Z_Q(p^2+M_Q^2)+O[(p^2+M_Q^2)^2]\,, \qquad
Z_Q = -E'_Q(-M_Q^2)>0 \,,
\end{eqnarray*}
where the positive constant $Z_Q$ defines the renormalization of the
wave function of the field $B_Q$. By using the standard
renormalization relation
$$\tilde{B}_Q(p)= Z_Q^{-1/2}\tilde{{\cal B}}_Q(p)$$
we write the kinetic term in the standard form:
\begin{eqnarray*}
&& \left(\tilde{B}_Q^+(p)\left[1-E_Q(-p^2)\right]\tilde{B}_Q(p)
\right)\\
&&= \left(\tilde{{\cal B}}_Q^+(p) \left[ (p^2+M_Q^2)
+O((p^2+M_Q^2)^2)\right]\tilde{{\cal B}}_Q(p)\right) \,.
\end{eqnarray*}

The interaction functional in (\ref{Z5}) reads
\begin{eqnarray}
\label{geff}
W_I[gB]=W_I[g_{\rm eff}{\cal B}] \,, \qquad
g_Q^{\rm eff} = g Z_Q^{-1/2}={g\over\sqrt{-E'_Q(-M_Q^2)}} \,>0.
\end{eqnarray}
It is essential, that according to (\ref{eigen}) the effective
coupling constant $g_Q^{\rm eff}$ does not depend on $g$ in any
explicit way.


\section{The Virton Model}

Let us consider a pure Gaussian case of analytic confinement
(\ref{prop1}), where the parameter $1/\Lambda$ represents the
"radius" of confinement. Due to the Gaussian character of these
propagators, the polarization kernel (\ref{kernel}) is quite simple
\begin{eqnarray}
\label{P}
\lambda\tilde{\Pi}_p(y,y')=\lambda{\Lambda^4\epsilon^2\over 2^6\pi^2}
\cdot e^{-{p^2\over 2\Lambda^2}}\cdot K(y,y')\,, \qquad
K(y,y')=e^{-{\Lambda^2\over 4}(y^2-yy'+y'^2)}
\end{eqnarray}
with $\lambda=\left({g / 4\pi\Lambda}\right)^2$. The kernel $K(x,y)$
can be explicitly diagonalized
\begin{eqnarray}
\label{K}
K(y,y')&=&\sum\limits_Q U_Q(y)\kappa_QU_Q(y'),\\
\kappa_Q &=& \kappa_{nl}=\kappa_0\cdot
\left({1\over2+\sqrt{3}}\right)^{2n+l},~~~~~~
\kappa_0=\left({2\pi\over\Lambda^2(2+\sqrt{3})}\right)^2 .\nonumber
\end{eqnarray}
The eigenfunctions $U_Q(y)$ are given in Appendix A.

According to (\ref{mass}) the mass spectrum of two-particle bound
states can be found explicitly
\begin{eqnarray}
\label{mass1}
M_Q^2&=&M_{nl}^2=M_0^2+(2n+l)\cdot2\Lambda^2\ln(2+\sqrt{3}) \,,\\
M_0^2&=& 2\Lambda^2\ln{\lambda_c\over \lambda},~~~~~~
\lambda_c=\left({4(2+\sqrt{3})^2\over\epsilon}\right)^2 \,.\nonumber
\end{eqnarray}

Thus, the Gaussian form of analytic confinement (\ref{prop1}) leads
to the purely linear and parallel RTs. The slope of RTs is defined
only by the scale of the confinement region $\Lambda$ and does not
depend on the coupling constant $\lambda$ and other dynamic constants
entering the interaction Lagrangian.

Bound states exist in the weak-coupling regime $\lambda<\lambda_c$.
If $\lambda\ll\lambda_c$ the size of the confinement region is
remarkably larger than the Compton length of all bound states
$$
r_{\rm conf}\sim{1\over\Lambda}\gg l_Q\sim{1\over M_Q}.
$$
In other words, the physical particles, described by fields $B_Q(x)$,
and all physical transformations involving them take place within the
confinement region.

\section{The Scalar Confinement Model}

The propagators of this model are given by (\ref{prop2}). In this case
the eigenvalue problem (\ref{eigen}) cannot be solved exactly, so a
variational approach will be used to evaluate approximately the
two-particle spectra. Further we consider only the orbital excitations,
i.e. the eigenvalues for zero radial quantum number $n=0$, so that
$Q=\{0,l,\{\mu\}\}$. If $\{\Psi_Q(y)\}=\{\Psi_{l\{\mu\}}(y)\}$ is a
set of normalized trial wave functions, then we have to calculate
\begin{eqnarray}
\label{vary}
&& \max\limits_{\Psi_Q}\sum\limits_{\{\mu\}}\int\!\!\!\!\int
dy_1dy_2~\Psi_Q(y_1)\lambda\Pi_p(y_1,y_2)\Psi_Q(y_2)
= \epsilon_l(M_l^2)\leq E_l(M_l^2)\nonumber\\
\end{eqnarray}
with $p^2=-M_l^2$. The mass of the bound state is determined by
the equation (\ref{mass}):
\begin{eqnarray}
\label{varmass}
&& \epsilon_l(M_l^2)=1.
\end{eqnarray}
Note, the solution of this equation gives the upper bound to the
mass $M_l^2$, because $\epsilon_l(M^2)\leq E_l(M^2)$ for positive
$M^2>0$.

In the present paper we choose the normalized trial wave function
in the following form
\begin{eqnarray}
\label{testf}
&& \Psi_{l\{\mu\}}(x,a)=C_l~T_{l\{\mu\}}(x)\sqrt{D(x)}~
e^{-{\Lambda^2\over4}ax^2},\\
&& C_l=\Lambda^{l+1}\sqrt{(1+2a)^{l+1} \over 2^l (l+1)!},~~~~~~~
\sum\limits_{\{\mu\}}\int dx \left| \Psi_{l\{\mu\}}(x)\right|^2 =1,
\nonumber
\end{eqnarray}
where $a$ is a variational parameter. The four-dimensional spherical
orthogonal harmonics $T_{l\{\mu\}}(x)$ are defined in Appendix A. We
guess that the test function in (\ref{testf}) should be a sufficiently
good approximation to the exact one because the kernel (\ref{kernel})
is proportional to $\sqrt{D(y)}$ and $S(y)$ is of the Gaussian type.

Let us define
\begin{eqnarray*}
\tilde{\Phi}_{l\{\mu\}}(k,a)&=&
i^l\int\!\! dx~e^{-ikx}\sqrt{D(x)}\Psi_{l\{\mu\}}(x,a)
={C_l\over (2\pi)^l}~T_{l\{\mu\}}(k)I(k^2),
\end{eqnarray*}
where
$$I(k^2)=\int\!\! d^{4+2l}Y e^{-iKY}D(Y)~e^{-aY^2}\,,~~~~~
K,Y\in{\bf R}^{4+2l},~~k^2=K^2.$$
Here the rotational symmetry $D=D(y^2)=D(Y^2)$ has been used. One
obtains
\begin{eqnarray}
\label{phiphi}
\sum\limits_{\mu}\tilde{\Phi}_{l\{\mu\}}(k,a)\,
\tilde{\Phi}_{l\{\mu\}}(k,a) &=& { C_l^2~k^{2l} (l+1) \over 2^{4+3l}}
\left[ \int\limits_{0}^{u_0} \!\! du u^l e^{-u k^2/4} \right]^2 \,,\\
u_0 &=& {4\over\Lambda^2(1+a)} \,.\nonumber
\end{eqnarray}

Substituting representations (\ref{prop2}), (\ref{testf}) and
(\ref{phiphi}) into (\ref{vary}) and after some calculations we
arrive at
\begin{eqnarray}
\label{energy}
\epsilon_l(M_l^2)&=&g^2\max\limits_a \!\int\!\! {dk\over (2\pi)^4}
\sum\limits_{\{\mu\}}\tilde{\Phi}_{l\{\mu\}}(k,a)\,
\tilde{G}\left(k+{p\over2}\right)\,\tilde{G}\left(k-{p\over2}\right)\,
\tilde{\Phi}_{l\{\mu\}}(k,a)\nonumber\\
&=&{2\lambda\over l!}\cdot\max\limits_c\left\{
[4c(1-c)]^{l+1}\int\!\!\!\!\int\limits_{0}^{1}\!\! dt ds \,
e^{-(t+s)(\mu^2-\kappa^2)}\,R_l(t,s,\chi)\right\}=1.\nonumber\\
\end{eqnarray}
Here
\begin{eqnarray*}
&& R_l(t,s,\chi)=\int\!\!\!\!\int\limits_{0}^{1}\!\! du dv~
e^{-\kappa^2\chi^2}~(uw)^lF_l(b,\chi),\\
&& F_l(b,\chi)=\int\!\! d^4k \,k^{2l}\, e^{-k^2 b-kp(t-s)}=
 e^{\chi^2/b}\left(-{\partial\over\partial b}\right)^l
\left[ {1\over b^2} \, e^{-\chi^2/b} \right]
\end{eqnarray*}

\begin{figure}[ht]
\begin{center}
\epsfig{figure=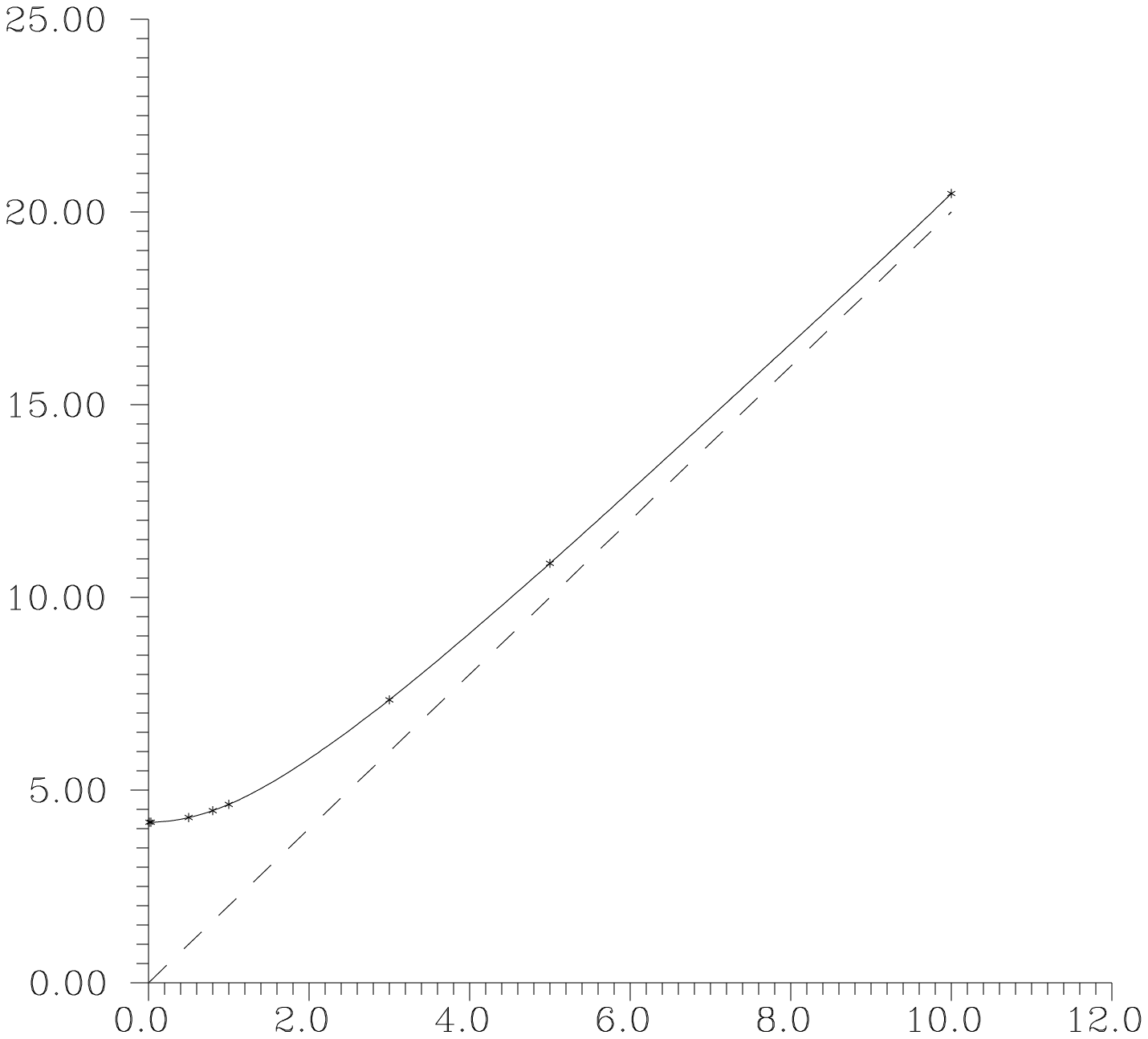,width=0.4\textwidth}

                         Fig. 1 \\[4mm]
\begin{minipage}{12cm}
{The spectrum of two-particle bound state $(M_0/\Lambda)$ at zero
orbital angular quantum number $l=0$ and $\lambda=0.01$ as function
of the mass $\mu=m/\Lambda$ of the "constituent" particle. The dashed
curve depicts the asymptotical line $2\mu$.}
\end{minipage}
\end{center}
\end{figure}

with $p=(iM,0,0,0)$ and $p^2=-M^2$, $b=t+s+2c(u+w)$ and
$$
\lambda=\left({g\over4\pi\Lambda}\right)^2,~~~~~\mu={m\over\Lambda},
~~~~~\kappa={M_l\over2\Lambda},~~~~~\chi^2=\kappa^2(t-s)^2 .
$$

The mass $M_l$ of the two-particle bound state in the one-parameter
variational approximation should be determined from the variational
equation (\ref{energy}). We have solved this equation numerically for
different values of parameters $\lambda$, $l$ and $\mu$. The obtained
results are plotted in Figs. 1-3.

\begin{figure}[ht]
\begin{center}
\epsfig{figure=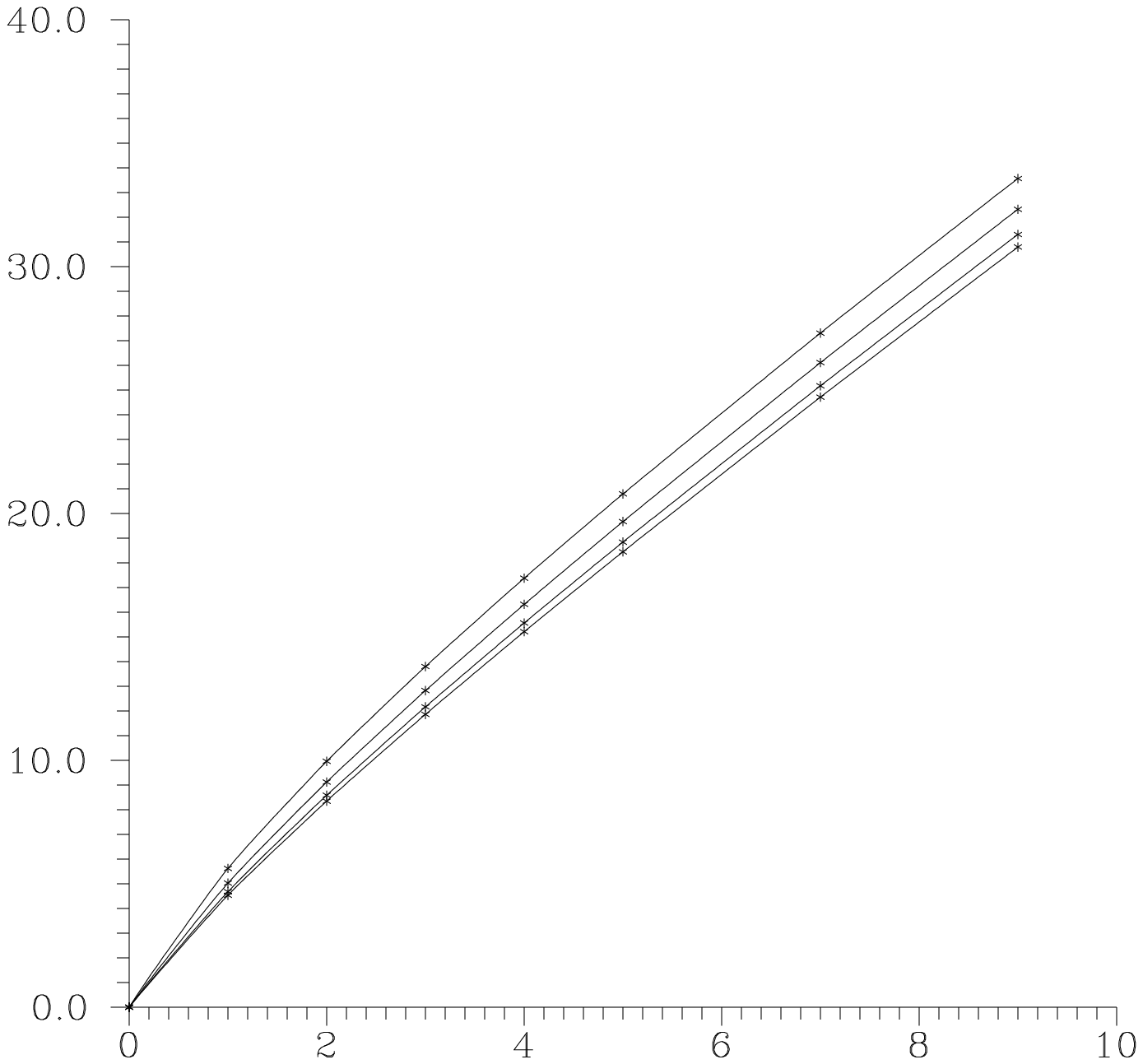,width=0.4\textwidth}

                         Fig. 2 \\[4mm]
\begin{minipage}{12cm}
{ The spectrum of two-particle bound state as function of the orbital
angular quantum number $l$ for several values of
$\mu=m/\Lambda=\{0.05;~0.2;~1.0;~5.0\}$ at fixed effective
coupling constant $\lambda=0.01$. For convenience the difference
$(M_l/\Lambda)^2-(M_0/\Lambda)^2$ is plotted.}
\end{minipage}
\end{center}
\end{figure}

\begin{figure}[ht]
\begin{center}
\epsfig{figure=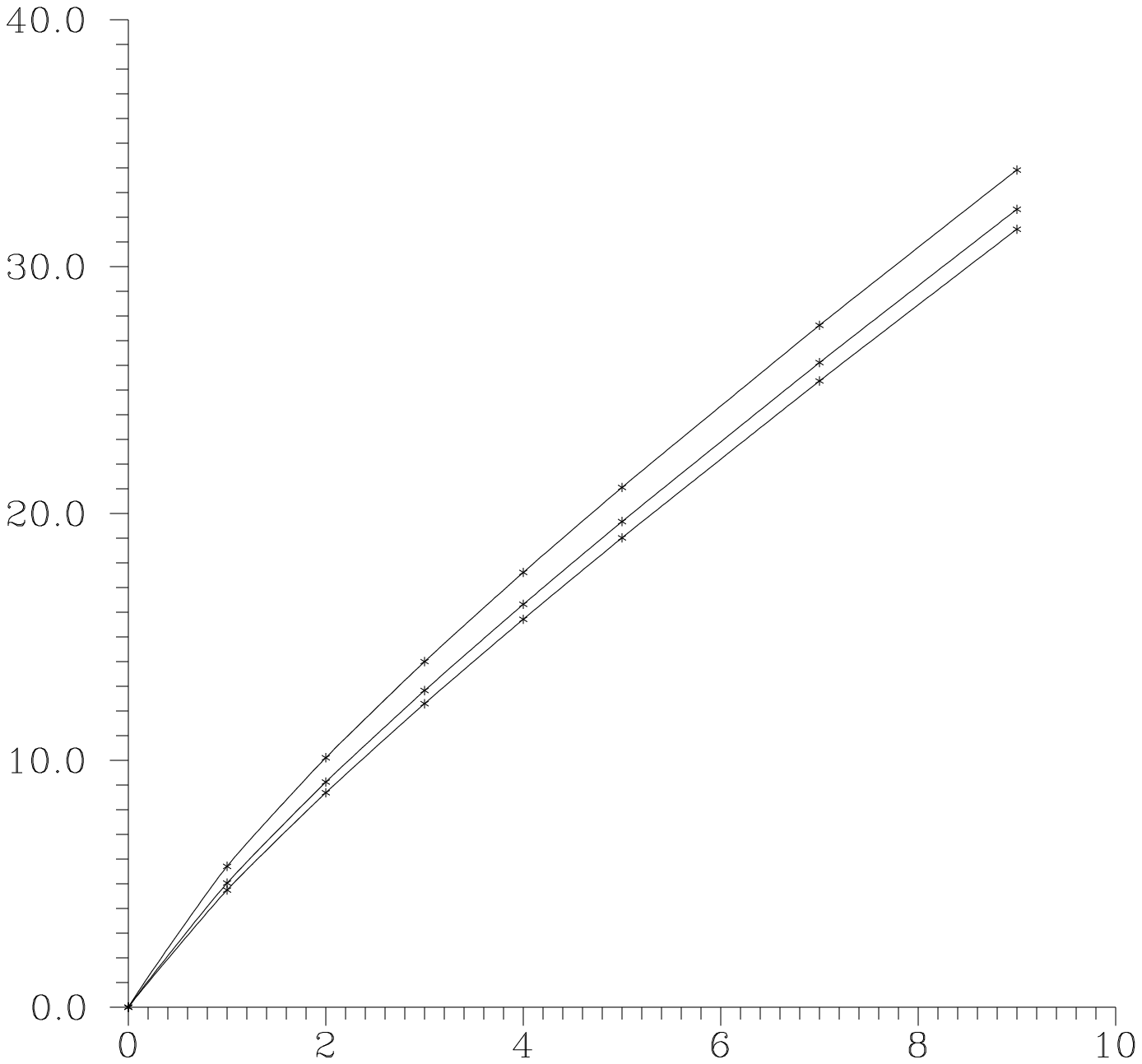,width=0.4\textwidth}

                         Fig. 3 \\[4mm]
\begin{minipage}{12cm}
{ The spectrum of two-particle bound state as function of the orbital
angular quantum number $l$ for several values of
$\lambda=\{0.001;~0.01;~0.1\}$ at fixed $\mu=1.0$. For convenience
the difference $(M_l/\Lambda)^2-(M_0/\Lambda)^2$ is plotted.}
\end{minipage}
\end{center}
\end{figure}


\section{Discussion}

Now let us make sure that the Scalar Confinement Model correctly
describes the basic features of the meson spectrum, when mesons are
considered bound states of quarks and gluons. In Fig.1 we show the
mass ${M_0/\Lambda}$ of the lowest bound state $B_0$ with quantum
numbers $Q=(0,0)=0$ as the function of the mass of constituent quarks
$\mu={m/\Lambda}$ for small coupling constant $\lambda=0.01$.

First of all, one can conclude that there exists a bound state in
the case $\mu=0$, i.e. with $m=0$. In other words, provided that the
analytic confinement takes place, two massless particles can be
coupled into a stable bound state (see also \cite{clpen98}). This
situation cannot be realized under any circumstances in the local
quantum field theory.

Second, the mass $M_0>2m$ for $\mu\leq 50\div 80$, i.e. it exceeds
the sum of masses of two constituent quarks. Besides, these bound
states are stable. It means that these heavy mesons are relativistic
systems and therefore, the nonrelativistic Schr\"{o}dinger equation
is not adequate tool to describe these heavy mesons. Probably the
masses of real heavy quarks (see Table 1) are not sufficiently "heavy"
to use nonrelativistic description. The conventional regime $M_0<2m$
appears for very large $\mu\geq 50\div 80$. In the deconfinement limit
$\Lambda\to 0$ our rough variational estimation (\ref{energy}) gives
the qualitatively correct behavior (for details see Appendix B)
\begin{eqnarray*}
&& M_0=2m-\left({g\over4\pi m}\right)^4 16\pi m C^2,
\end{eqnarray*}
i.e. we get the standard nonrelativistic (the coupling constant
$g/4\pi m$ is small!) bound state under the Coulomb potential.

\begin{center}
\begin{tabular}{|c|c||c|c|}
\hline
 & & & \\
Quarks & Mass ({\it Mev}) & Meson & Mass ({\it Mev})\\
 & & & \\
\hline
 & & & \\
$u$ & 1$\div$5 & $\rho=u\bar{u}$& 770\\
$d$ & 3$\div$9 & & \\
$s$ & 75$\div$170 & $K^*=s\bar{s}$ & 900\\
$c$ & 1150$\div$1350 & $\eta_c=c\bar{c}$ & 2980\\
$b$ & 4000$\div$4400 & $\gamma=b\bar{b}$ & 9460\\
 & & & \\
\hline
\end{tabular}
\vskip 4mm

                    Table 1. \\[4mm]
\begin{minipage}{12cm}
{ The experimentally observed masses of light and heavy quarks $q$
and mesons $q\bar{q}$ \cite{PDG}.}
\end{minipage}
\end{center}

In  Table 1 we list the masses of quarks and mesons as bound states
of corresponding quarks. One can see that the meson masses are larger
than two masses of the constituent quarks. The Scalar Confinement
Model discribes this property of meson spectrum.

In Figures 2 and 3 we plot the Regge trajectories, i.e  the squared
masses of orbital excitations at different values of the small
coupling constant $\lambda$ and masses of constituent quarks $\mu$.
One can see that the dependence of RTs on the parameters $\lambda$
and $\mu$ is not strong. Besides, the RTs are not linear, although
the linearity  begins in fact for $l\geq 4\div 6$. The asymptotical
behavior of the Regge spectrum for sufficiently large $l$ can be
obtained analytically and coincides with the exact solution of the
Virton Model (\ref{mass1}):
\begin{eqnarray}
\label{asym}
M^2_l\sim l\cdot\Lambda^22\ln(2+\sqrt{3})~~~~{\rm for}~~~~l\to\infty.
\end{eqnarray}

On the other hand, the dependence $M_l^2$ on $l$ is quite smooth, so
that three or four lowest orbital excitations can be approximatly
considered almost linear. Therefore, they can be well approximated by
linear trajectories on a short interval $l=0 \div 3$.

A recent analysis of the experimental data shows (see \cite{tang00})
that the RTs of different meson and baryon families are approximately
linear and their slopes deviate slightly around a constant value,
although the quark configurations and quantum numbers of these
hadronic families are considerably different. Note, the analysed
data in \cite{tang00} have been obtained for low orbital momenta
$l=0\div 3$ only.

Nevertheless, one can conclude that the slope of RTs weakly depends
on specific details of hadron internal dynamics and may be considered
an universal characteristic which is dictated only by the general
properties of quark-gluon interactions.

Precisely this qualitative picture takes place for our models with
analytic confinement. Thus, we have sufficient ground to claim that
the analytic confinement realizes these general properties and leads
to the approximate linearity of RTs for meson families.

Therefore, the Scalar Confinement Model should qualitatively describe
the correct behavior of orbital excitations. We can compare the
experimental data with our calculations. In \cite{tang00} the RTs of
$\pi$-, $K$-, $\rho$-, $\omega$-, $\phi$- and $K^*$-meson families
have been fitted on the interval $0\leq l\leq 3$ by curves
$M^2_l=m_0+m_1l+m_2l^2$ with appropriate constants $m_j$. We have also
performed fittings of our data obtained for $l=1,2,3$. By comparing
corresponding coefficients of fittings for $\pi$- and $K$-meson
trajectories with our fit coefficients obtained at different values of
parameters $\mu$ and $\lambda$ we have found the possible values of
the confinement scale $\Lambda$. The result is given in Table 2. One
can see that $\Lambda$ depends weakly on $\mu$ and $\lambda$ in wide
ranges. Our semiquantative result indicates that the confinement scale
parameter is around $\Lambda_{conf}\approx 500~{\it MeV}$.

\begin{center}
\begin{tabular}{|c|l|c|c|}
\hline
& & & \\
$\mu$ & $\lambda$ & $\Lambda_{\pi}~({\it MeV})$ & $\Lambda_K~({\it MeV})$ \\
& & & \\
\hline\hline
& & & \\
0.2 & 0.01 & 562 & 504 \\
1.0 & 0.001 & 558 & 502 \\
1.0 & 0.01 & 539 & 487 \\
1.0 & 0.1 & 502 & 452 \\
5.0 & 0.01 & 506 & 456 \\
& & & \\
\hline
\end{tabular}
\vskip 4mm

                    Table 2. \\[4mm]
\begin{minipage}{12cm}
{ The scale of confinement $\Lambda$ calculated from the comparison
with real spectra of $\pi$- and $K$-meson families given in
\cite{tang00}.}
\end{minipage}
\end{center}

\acknowledgments

The authors would like to thank I.Ya.~Aref'eva, V.Ya.~Fainberg and
A.A.~Slavnov for useful discussions.

\vskip 1cm
{\bf Appendix A} \\[3mm]

Consider the kernel
\begin{eqnarray}
\label{AK}
K=K(x,y)=e^{-ax^2+2bxy-ay^2} \,, \quad a>b
\end{eqnarray}
with
\begin{eqnarray*}
{\rm Tr}~K = \int dy~K(y,y)=\int dy~e^{-2(a-b)y^2} =
 {\pi^2\over4(a-b)^2}<\infty \,.
\end{eqnarray*}
The eigenvalues with quantum numbers $Q=\{nl\{\mu\}\}=
\{nl\{\mu_1...\mu_l\}\}$ and eigenfunctions of the problem
\begin{eqnarray*}
\int dy~K(x,y)U_Q(y) = \kappa_QU_Q(x)
\end{eqnarray*}
can be solved explicitly. The eigenvalues are
\begin{eqnarray}
\label{AV}
&& \kappa_Q=\kappa_{nl}=\kappa_0\cdot
\left({b\over a+\sqrt{a^2-b^2}}\right)^{2n+l} \,, \qquad
\kappa_0={\pi^2\over(a+\sqrt{a^2-b^2})^2} \,.
\end{eqnarray}
The eigenfunctions are
\begin{eqnarray}
\label{AU}
U_Q = U_{nl\{\mu\}}(y) = N_{nl} \, T_{l\{\mu\}}(y) \,
L_n^{(l+1)}\left(2\beta y^2\right)e^{-\beta y^2} \,.
\end{eqnarray}
Here $L_n^{(l+1)}(x)$ are the Laguerre polynomials and
\begin{eqnarray*}
\beta=\sqrt{a^2-b^2} \,, \quad
N_{nl}={\sqrt{2^l(l+1)}\over\pi}\cdot(2\beta)^{1+{l\over2}}\cdot
\sqrt{{\Gamma(n+1)\over\Gamma(n+l+2)}}
\end{eqnarray*}
The functions
$$
T_{l\{\mu\}}(y)=T_{l\{\mu\}}(n_y)|y|^l,~~~~~~n_y={y\over|y|},
~~~~~~ |y|=\sqrt{y^2}
$$
satisfy the conditions
$$
T_{l\{\mu_1\mu_2...\mu_l\}}(n)=T_{l\{\mu_2\mu_1...\mu_l\}}(n)\,,
\qquad T_{l\{\mu\mu\mu_3...\mu_l\}}(n)=0 \,,
$$
\begin{eqnarray*}
\sum\limits_{\{\mu\}}T_{l\{\mu\}}(n_1) \,  T_{l\{\mu\}}(n_2)
= {1\over2^l}C_l^1((n_1n_2)),~~~~~~~C_l^1(1)=l+1 \,,
\end{eqnarray*}
where $C_l^1(t)$ are the Gegenbauer polynomials and
\begin{eqnarray*}
\int\!\! dn~T_{l\{\mu\}}(n) \, T_{l'\{\mu'\}}(n)
=\delta_{ll'}\delta_{\{\mu\}\{\mu'\}}\cdot{2\pi^2\over2^l(l+1)} \,.
\end{eqnarray*}

Besides, the following relation takes place
\begin{eqnarray}
\label{equa1}
&& \int d^4 y~T_{l\{\mu\}}(y) F(y^2) e^{-iky} \,
=(-i)^l~T_{l\{\mu\}}(k) J(k^2),               \\
&& J(k^2)=\int dY e^{-iKY}~F(Y^2),~~~~K,Y\in{\bf R}^{4+2l},
~~~~~ k^2=K^2\,.    \nonumber
\end{eqnarray}

\vskip 1cm
{\bf Appendix B} \\[3mm]

Consider the variational problem (\ref{energy}) in the deconfinement
limit $\Lambda\to 0$ for $l=0$
\begin{eqnarray}
\label{regge3}
&& 8\left({m\over\Lambda}\right)^2\lambda_0\max\limits_{0<c<1}
\left\{c(1-c)\int\!\!\!\!\int\limits_{0}^{1} dt ds~e^{-\left(
{m^2\over\Lambda^2}-{M_0^2\over4\Lambda^2}\right)(t+s)}\right.\\
&&~~~~~~~~~~~~\left.\cdot\int\!\!\!\!\int\limits_{0}^{1}\!\! du du~
{\exp\left\{-{M_0^2\over 4\Lambda^2}~{(t-s)^2\over t+s+2c(u+v)}
\right\} \over [t+s+2c(u+v)]^2 } \right\}=1 \,. \nonumber
\end{eqnarray}
Here $M_0$ is the mass of the lowest bound state and the effective
coupling constant is given as
\begin{eqnarray*}
\lambda_0=\left({g\over4\pi m}\right)^2\ll1 \,.
\end{eqnarray*}
Going to the new variables
$$
t = {\Lambda^2\over 2m^2}(x+y) \,, \qquad
s = {\Lambda^2\over 2m^2}(x-y) \,, \qquad
c = {\Lambda^2\over m^2} \xi
$$
we rewrite (\ref{regge3})
\begin{eqnarray*}
\label{regge4}
&& 4\lambda_0 \max\limits_\xi\left\{ \xi \left(
1-{\Lambda^2\over m^2} \xi\right)
\int\limits_{0}^{{m^2\over\Lambda^2}}\!\! dx
~e^{-\left(1-{M_0^2\over 4m^2}\right)x}\right.\\
&&~~~~~~~~~~~~~~~~\left.\cdot\int\!\!\!\!\int\limits_{0}^{1}\!\!
{dudv\over[x+2\xi (u+v)]^2}\int\limits_{-x}^{x}\!\! dy
~e^{-{M_0^2\over 4m^2}~{y^2\over x+2\xi (u+v)}} \right\}=1
\,.\nonumber
\end{eqnarray*}
Obviously, the limit $\Lambda\to 0$ exists if $M_0<2m$. Besides, since
$\lambda_0\ll 1$ and $1-{M_0\over 2m}\ll 1$, the main contribution
into the integral over $dx$ comes from large $x$, so that the inner
integral over $dy$ can be explicitly taken on the extended interval
$\{-\infty,\infty\}$ without any loss of accuracy. Thus, we get
\begin{eqnarray}
\label{estim1}
&& {8m\lambda_0 \over M_0}
\sqrt{{\pi\over 1-M_0^2/4m^2}}\cdot C =1 \,,\\
&& C = \max\limits_{0<\xi<\infty}\left\{
\xi \int\limits_{0}^{\infty}\!\! dx ~e^{-x}
\int\!\!\!\!\int\limits_{0}^{1}\!\!
{du dv \over [x+2\xi (u+v)]^{3/2}} \right\}=0.31923... .\nonumber
\end{eqnarray}
By solving (\ref{estim1}) we obtain the mass of the lowest
two-particle bound state in the deconfinement limit $\Lambda\to 0$
as follows
\begin{equation}
\label{coulomb}
M_0=2m - \lambda_0^2 \cdot 16\pi m C^2 + O(\lambda_0^4)\,.
\end{equation}
This behavior is in accordance with the conventional nonrelativistic
picture of the attractive Coulomb interaction when the binding energy
of two-particle bound system is negative.


\end{document}